\documentclass[%
 reprint,
superscriptaddress,
showpacs,preprintnumbers,
amsmath,amssymb,
aps,
prper,
]{revtex4-2}

\usepackage{graphicx}
\usepackage{dcolumn}
\usepackage{bm}
\usepackage{hyperref}

\begin{document}

\title{Quantum mechanics curriculum in the US: Quantifying the instructional time, content taught, and paradigms used}

\author{Alexis Buzzell}
\author{Ram\'on Barthelemy}
\affiliation{%
 Department of Physics \& Astronomy, University of Utah
}%

\author{Tim Atherton}%
\affiliation{%
 Department of Physics \& Astronomy, Tufts University
}%

\begin{abstract}
 Quantum mechanics is an integral course for physics students. An understanding of quantum concepts is imperative for enrollment in physics graduate programs, participating in research within physics-fields, and employment at companies developing quantum technologies. This study analyzes 188 US research intensive institutions' course catalogs to determine the role and volume of quantum mechanics in their undergraduate physics programs. All of the institutions required one course on quantum concepts, 92\% required two courses, and half required three. For institutions with complete class data ($n=56$), the  quantum curriculum was analyzed using course syllabi. The mean number of classroom hours spent on quantum concepts was found to be 63.5 hours with a standard deviation of 28.1 hours. The most commonly taught themes in the quantum curriculum were the Schr{\"o}dinger equation and three-dimensional quantum mechanics. However, the Stern-Gerlach Experiment was only included in 28\% of the course outlines. Despite current efforts to promote a spin-first approach, this study found the traditional position-first approaches were still more common as they were used by 73.7\% of instructors. 
\end{abstract}

\maketitle

\section{Introduction}
Popularized by Gladwell's \textit{Outliers: The Story of Success}, there is a common notion that it takes 10,000 hours of practice at a skill to become an expert \cite{Gladwell}. While this idea is an oversimplification, as Gladwell himself and others have pointed out, factors such as family, culture, and friendship play a role in success \cite{Gladwell,Wong}. While time is not a perfect true indicator of expertise, it is one of the many factors that are important. As the quantum workforce grows, the field of physics must determine how much time institutions should devote in their physics curriculum to quantum concepts. 

Institutions have popularized the notion that for every one hour spent in lecture or class, students should expect to spend two to three hours of their own time studying \cite{Time1, Time2, Time3, Time4, Time5}. This number may be larger or smaller depending on the individuals' abilities and outside commitments, but it does provide a relative gauge on  the number of hours students are expected to spend on a topic during their four year degree based on the scheduled lecture times of their courses. As physics educators, we may want to know how much time is necessary for a student to be considered an expert and prepared to work within the quantum workforce. As outlined by the US National Quantum Initiative in the Quantum Information Science and Technology Workforce Development National Strategic Plan, current goals for the quantum workforce are to assess gaps in education and training opportunities, as well as make the quantum sector more accessible and equitable \cite{NationalQMStrategy}. This study will support this effort by determining the current state of US physics programs' quantum educational training, as well as assessing if there is equal access to educational opportunities across institutions. 

Building off previous work of a modern physics syllabi analysis, this paper will be the first to quantify the current quantum mechanics curriculum of research intensive institutions in the US. This study will answer the following research questions:
\begin{enumerate}
    \item \textit{How many courses on quantum concepts are physics students required to complete to be awarded a four year degree?}
    \item \textit{How many classroom hours on quantum phenomena are physics majors required to take in order to graduate with a four-year degree in physics?}
    \item \textit{What topics taught in quantum mechanics are required for students to learn to be awarded a four-year degree in physics? }
    \item \textit{Are institutions using a spin-first (using linear algebra) or a position-first (using differential equations) pedagogical approach?}
\end{enumerate}

\section{Literature review}
Previous work found that the modern physics course is most often students' first introduction to quantum concepts within the US physics curriculum. However, with the most common prerequisite for modern physics being calculus II, it was concluded that physics students would need to enroll in another course on quantum mechanics in order to be able to solve the Schr{\"o}dinger equation at the level of Griffiths' \cite{Griffiths}, a commonly used undergraduate textbook. In addition, work by Singh has suggested that quantum mechanics education in physics programs may not be satisfactory for career preparedness in quantum related fields (e.g. quantum computing) \cite{Singh2007}. The conclusions of the modern physics course analysis, along with Singh's findings, motivated this study to characterize the quantum curriculum across all four years of undergraduate physics programs in the US. The quantification of the current quantum curriculum should be the start of an ongoing conversation in the physics community to determine if physics programs are preparing students for careers in quantum related fields, as well as ensuring students at all US institutions have access to a comprehensive quantum education. 

While no study has previously characterized whether or not US institutions are meeting a list of common goals in their quantum mechanics courses, a survey of thirteen quantum mechanics instructors found they all shared common objectives for their courses \cite{SiddquiSingh2010}. These thirteen instructors all shared that they want their students to learn the postulates of quantum mechanics and the formalisms, and that they can apply these formalisms to solve a variety of different problems.

Within the physics education research community, there have been many efforts to achieve these commonly shared goals and increase student learning within the abstract realm of quantum mechanics. One such effort has been the commonly debated spin-first versus position-first approach of teaching quantum mechanics. Within a position-first paradigm, students are introduced to the Schr{\"o}dinger Equation early on in the course \cite{Passante, Sadaghiani}, and it is introduced within the context of wave functions in position space \cite{Passante}. Differential equations are the primary mathematical tool in understanding these equations, and the Time Independent Schr{\"o}dinger Equation is the first eigenvalue equation introduced within the course. "The focus is more on carrying out mathematical calculations rather than sense making from experimental results" \cite{Sadaghiani}. A typical textbook utilized in these courses is Griffiths' \textit{Introduction to Quantum Mechanics} \cite{Griffiths, Passante}. 

However, in a spin-first paradigm the Stern-Gerlach Experiment (SGE) is often how quantum concepts are introduced to students \cite{Passante}. The postulates of quantum mechanics are usually explored through the SGE and explicitly mentioned \cite{Passante, Sadaghiani}. Within this paradigm, students are introduced to the Schr{\"o}dinger Equation through the context of spin-1/2 particles, and have regularly used eigenvalue equations in the context of quantum mechanics before the Schr{\"o}dinger Equation is introduced \cite{Passante}. Students utilize matrix equations and linear algebra as their primary mathematical tool to understand the phenomena. McIntyre's \textit{Quantum Mechanics: A Paradigms Approach} is a commonly used textbook \cite{Passante, McIntyre}. 

Some of the supporters of the spin-first paradigm argue that this approach can help students overcome the commonly found difficulty in discerning physical space and Hilbert space that occurs through position-first approaches \cite{Passante, Sadaghiani}. Singh was able to illuminate this difficulty in a previous study where it was found only 41\% of students were able to sketch the physical results on the detector of the SGE, revealing their difficulty with physical space versus Hilbert space \cite{Singh2007185}. However, after the development of the SGE Quantum Interactive Learning Tutorial 
(QuILT), 22 students in a course were able to improve their pretest scores from a 53\% average to a 92\% average on the post-test scores \cite{ZhuSingh2009}. This finding confirmed the potential a spin-first paradigm could offer students in combating this common misconception. Singh has also hypothesized that without the knowledge of wave function collapse, linear algebra students may be able to answer similar questions correctly more often than students within a quantum mechanics course in a position-first paradigm \cite{Singh2007}. 

To compare the two paradigms, Sadaghiani \cite{Sadaghiani} had two course sections of sophomore level modern physics taught by the same instructor at Cal Poly Pomona, one where spin-first paradigm was used and the other where position-first paradigm was used. It was concluded that the spin-first students outperformed the position-first students with average scores of 63\% vs 57\% and 55\%. The survey results found that students in the spin-first approach had shifted their focus from computation to sense-making by providing concrete experimental evidence and simplifying the mathematical calculations involved. 

Other scholars have also studied a spin-first paradigm \cite{Passante}. Using a resources lens, Passante considered the tools and methods students instructed through each paradigm used to solve questions related to energy measurement. Specifically, Passante is interested in how energy and the Hamiltonian are introduced within each paradigm and the result of how students conceptualize energy measurement. Both paradigms introduce energy as the eigenvalue of the Hamiltonian operator to the Schr{\"o}dinger Equation, but the form the Hamiltonian takes differs in the two paradigms. Passante found that the most common resources students used in the spin-first paradigm were: (1) energy is an eigenvalue (43\%), (2) probability is calculated from \begin{math} |\langle \psi | \psi \rangle |^2 \end{math} (38\%), and  (3) the operator associated with energy is the Hamiltonian (38\%). The resources used by students in the position-first paradigm were most commonly: (1)  multiple values are possible for a probability measurement of a superposition state (33\%), (2) use of the expectation value (17\%), and (3) coefficients in the state are meaningful (13\%). Passante was able to come to several preliminary findings, such as students in the position-first paradigm activated resources related to the role of imaginary numbers and expectation values that the spin-first students did not activate, and students in the spin-first paradigm may form a more pronounced connection between the Hamiltonian and energy. Other work has focused on faculty perspectives, rather than student progress.

Siddiqui and Singh \cite{SiddquiSingh2010} uncovered what student difficulties instructors  see most often in quantum mechanics courses. The instructors commented on the inaccurate descriptions the students learn in modern physics courses, which were exacerbating their difficulties. One instructor noted that the misconceptions in quantum mechanics are avoidable if we do not teach students inaccurate descriptions of quantum systems that we commonly use in the modern physics courses. One instructor stated, "I find the Bohr model hideous. I really wish modern physics classes could go beyond the Bohr model some day, so that we don't have to unteach it later" \cite{SiddquiSingh2010}. Two thirds of the instructors shared this belief. This argument, combined with the arguments for a spin-first paradigm, suggests that a position-first traditional education in modern physics courses is one component that could contribute to a student's lack of understanding of  quantum phenomena. This finding motivated this study to characterize what paradigms institutions are using in their courses. 

\section{Methods}
The US News Rankings of "The Best Physics Programs" was used as an institution reference list in this study \cite{USNEWS}.  This list ranks mostly research focused physics programs that offer graduate level physics education. We chose to focus on this community of schools as they are producing the next generation of physics faculty and have been shown to educate the majority of students in a physics education \cite{AIP}. This list ranks 190 institutions, of which 188 offer a four-year physics degree. In the year 2021-2022, these institutions cumulatively granted 4,772 bachelor's degrees, which is 56.7\% of the bachelors degrees in physics granted in that year. This same list was utilized in our previous study on modern physics course syllabi.

Using this reference list of 188 institutions, the number of courses on quantum concepts required for a four-year degree in physics was determined by referencing course catalogs publicly available online. These 188 institutions are 74.5\% very high research activity, and 22.3\% high research activity in the Carnegie Classification of Higher Education Institutions \cite{Carnegie}, with 6 institutions research activity not defined by these metrics. 69.7\% of the institutions are public, and 30.3\% are private institutions \cite{Carnegie}. 16.5\% are Minority Serving Institutions (MSI, $n=31$), including Hispanic Serving Institutions (HSI, $n=17$), Asian American and Native American Pacific Islander-Serving Institutions (AANAPISI, $n=13$), Historically Black Colleges and Universities (HBCU, $n=3$), Predominantly Black Institutions (PBI, $n=1$), and Alaska Native-Serving Institutions or Native Hawaiian-Serving Institutions (ANNH, $n=2$) \cite{MSI}. We understand the uniqueness of MSIs and wanted to characterize their quantum curriculum separately. To do so, the MSI's required course data was used in the larger data set, but also went through a round of analysis separately.

Using the institutions' course catalogs, the requirements for a four year degree in physics were referenced. All required course descriptions were analyzed and coded as either including quantum concepts or not including quantum concepts. Any required course that included quantum topics, "modern physics," or "modern topics" was recorded as a quantum related course, as a previous finding showed that quantum is the most commonly taught topic within modern physics courses. 

If the institution offered different concentrations for students to choose from that changed their degree requirements, the most generalized degree option was selected for coding. For example, if a school offered a bachelor of arts degree or a bachelors of science degree, the bachelors of science was selected for coding. If the institution offered an applied physics degree along with other concentrations, the applied physics degree was selected for coding.

To determine how many classroom hours are scheduled during the required courses, syllabi had to be collected, specifically for courses such as modern physics or courses with "modern topics" in their course description. In order to count the hours spent on quantum concepts in courses that include topics other than quantum physics (such as relativity, or thermodynamics), a detailed course schedule was required to be able to discern how many lectures were spent on quantum topics compared to the other topics. 

Utilizing the modern physics syllabi collected in a previous study made this process easier for the introductory courses. If the syllabi needed had not already been collected from this previous study, an online search was first utilized to see if the syllabus was available publicly. If the syllabus was not, then an online course schedule was used to find a recent instructor of the course. An email was sent to the instructor requesting a copy of their syllabus and course schedule. 

This syllabi collection process resulted in a total of 300 syllabi across the 188 institutions. However, only 56 institutions had syllabi which specified which lectures covered quantum topics and what lectures covered relativity, thermodynamics, or other topics in their introductory courses, so they are the only ones included in this portion of the analysis. From these 56 institutions, a total of 129 syllabi were collected, 51.2\% from public sources and 48.8\% through private correspondence. The 56 institutions awarded 1,746 physics bachelors degrees in the year 2021-22 \cite{AIP} or 20.8\% of physics bachelors in that year. 69.6\% are public universities, 82.1\% are considered to have very high research activity, and 16.1\% have high research activity \cite{Carnegie}. 16.1\% of these institutions are MSIs \cite{MSI}.

In order for a lecture to be coded as a quantum related lecture, the same codebook was used as in the previous modern physics study. The lecture had to include one of the words listed in Table \ref{tab:quantumlecture} in order to be coded as a "quantum related lecture."

\begin{table}
\centering
\begin{tabular}{ |c|c| }
    \hline
    \multicolumn{2}{|c|}{Codes to be considered a quantum lecture} \\
    \hline
         Schr{\"o}dinger& Fermi's golden rule  \\
         Schr{\"o}dinger equation& Photons\\
         Photoelectric effect& Pauli's exclusion principle\\
         Wave-particle duality& Square well\\
         Operators& Identical particles\\
         Eigenvalues& Matter waves\\
         Tunneling/reflection& Frank Hertz experiment\\
         Stern-Gerlach experiment& Wave mechanics\\
         Dirac Notation& Wave functions\\
         States& Wave properties of particles\\
         Quantum measurement& Particle properties of waves\\
         Expectation value& de Broglie Hypothesis\\
         Uncertainty& Quantum theory of light\\
         Superposition& Blackbody radiation\\
         Mixed states& Planck's postulate\\
         Quantization& Spin\\
         \hline
    \end{tabular}
    \caption{Codes for a lecture to be considered a quantum related lecture}
    \label{tab:quantumlecture}
\end{table}

    \begin{table*}
    \centering
    \begin{tabular}{ |c|c|c| }
    \hline
    \multicolumn{3}{|c|}{Themes and related topics present in the syllabi} \\
    \hline
       \textbf{Quantum theory of light}& \textbf{Schr{\"o}dinger equation}& \textbf{Particles} \\
       \hline
       Photons& Time independent Sch{\"o}dinger equation& Pauli exclusion principle\\
       Compton effect& 1D quantum mechanics& Elementary particles\\
       Wave-particle duality& Solutions to different potentials& Fermions\\
       Particle nature of matter& Potential well(infinite and finite)& Bosons\\
       Wave nature of particles& Harmonic oscillator& Identical particles\\
       & Bound states& \\
       \hline
       \textbf{Wave functions}& \textbf{Scattering theory}& \textbf{EPR Paradox}\\
       \hline
       Wave mechanics& Born approximation& Bell's theorem\\
       \hline
       \textbf{Applications}& \textbf{Tunneling}& \textbf{Adiabatic approximation}\\
       \hline
       Quantum computing& Finite square well& Adiabatic theorem\\
       Quantum teleportation& Reflection/scattering& Berry's phase\\
       Cryptography& Unbound states& \\
       \hline
       \textbf{3D quantum mechanics}& \textbf{Formalism}& \textbf{Early quantum experiments}\\
       \hline
       Spherical coordinates& Operators& de Broglie wavelengths\\
       Hydrogen atom (in context& Observables& Blackbody radiation\\
       (of Sch{\"o}dinger equation,& Measurables& Planck's hypothesis\\
       no Bohr model)& Expectation value& Photoelectric effect\\
       Angular momentum& Hilbert space& Double slit experiment\\
       & Eigenvalues/functions& Frank Hertz experiment\\
       & Normalization& \\
       & Probability& \\
       \hline
       \textbf{Perturbation theory}& \textbf{Time evolution}& \\
       \hline
       Time independent & Time development& \\
       perturbation theory& Quantum dynamics& \\
       Time dependent& Spin procession& \\
       perturbation theory& Lamor procession& \\
       Nondegenerate& Neutrino oscillations& \\
       perturbation theory& Time dependent& \\
       Degenerate perturbation& Hamiltonians& \\
       theory& & \\
       Fine structure& & \\
       Zeeman effect& & \\
       Hyperfine splitting& & \\
       Spontaneous emission& & \\
       \hline
    \end{tabular}
    \caption{Topics combined into overall themes for coding syllabi}
    \label{tab:themesandtopics}
    \end{table*}

Using this coding scheme, the number of quantum lectures were tallied up for the course. If the course was not an introductory course and was instead a traditional quantum mechanics course, i.e. only covers quantum concepts through the entire semester, the institution's publicly available academic calendar was consulted to determine the number of lectures the course held during the semester. The scheduled lecture time was then used to discern the number of minutes within each lecture. From here the total number of quantum related lecture minutes was calculated by multiplying the number of quantum lectures by the number of minutes per lecture.

To ensure both methods of counting (syllabus schedule vs. academic calendar) midterm exams held after quantum concepts were introduced were included in the calculation. Recitations or discussion sections were not included in the count, as not all courses had recitation or discussion sections, and these may be considered part of the time students spend studying outside of lecture time as these sections are usually reserved for practice problems and group work. 

In order to characterize the topics taught within the quantum curriculum, the 56 institutions for which all quantum related course syllabi were used for analysis. Restricting analysis to these 56 institutions ensured that the entire quantum curriculum for each institution was characterized rather than only one or two classes being characterized. From these 56 institutions, only 50 were able to be analyzed for their topics, as one or more syllabi from six of the institutions did not list out the topics the course would cover. 

To determine the topics taught, an emergent coding method was used \cite{Creswell}. All topics mentioned in the syllabi were recorded. Using the table of contents within the Griffiths' \cite{Griffiths} and McIntyre \cite{McIntyre} textbooks, topics that were closely related combined together. Table \ref{tab:themesandtopics} below presents the overarching themes within the syllabi and topics that are coded as being within each theme. Other topics appeared so few times, no other topics were grouped together with it and will appear in the results despite not being in Table \ref{tab:themesandtopics}.

\begin{table*}
    \centering
    \begin{tabular}{ |c|c| }
    \hline
    \multicolumn{2}{|c|}{Coding for paradigm used by course}\\
    \hline
    \textbf{Spin-first}& \textbf{Position-first}\\
    \hline
    Stern-Gerlach experiment \cite{Passante}& Schr{\"o}dinger equation introduced\\
    Postulates of quantum mechanics \cite{Passante, Sadaghiani}& early on \cite{Passante, Sadaghiani}\\
    Schr{\"o}dinger equation in context& Schr{\"o}dinger equation in context\\
    of spin 1/2 particles \cite{Passante}& of position space wave functions \cite{Passante}\\
    Matrix equations \cite{Passante}& Differential equations \cite{Passante} \\
    Eigenvalue equations regularly& Time independent Schr{\"o}dinger equation\\
    used before Schr{\"o}dinger equation introduced \cite{Passante}& is first eigenvalue equation introduced \cite{Passante}\\
    McIntyre's textbook used \cite{Passante, McIntyre}& Griffiths' textbook used \cite{Passante, Griffiths}\\
    \hline
    \end{tabular}
    \caption{Codes to determine paradigm used by each course}
    \label{tab:spinposition}
\end{table*}
    
Lastly, the paradigm institutions were using (spin or position first) was analyzed twice: once for institutions' introduction to quantum courses (i.e. modern physics courses), and another time for institutions' traditional quantum mechanics courses. The rationale for this step was due to a previous study that found the most common prerequisite for modern physics was calculus II. Without a foundation in linear algebra, the question was posed on whether or not students were prepared for a spins first approach. Therefore, this study decided to determine if any institutions were utilizing a spins first approach in their modern physics or equivalent course.

For the modern physics or equivalent courses, 112 syllabi from different institutions were collected and analyzed. These 112 institutions awarded 2,785 physics bachelors degrees (33.1\%) in 2021-22 \cite{AIP}, 74.1\% are public, 75.0\% have very high research activity, and 21.4\% have high research activity \cite{Carnegie}. 16.1\% are MSIs, with HSI ($n=10$), AANAPISI ($n=7$), and ANNH ($n=1$) represented \cite{MSI}. 

For the quantum mechanics courses, 99 syllabi were collected and analyzed. From these 99 institutions, 2,920 physics bachelors degrees (34.7\%) were awarded in 2021-22 \cite{AIP}, 70.7\% are public, 79.8\% have very high research activity, and 17.1\% have high research activity \cite{Carnegie}. 17.2\% are MSIs with HSI ($n=10$), AANAPISI ($n=7$), HBCU ($n=1$), and ANNH ($n=1$) represented \cite{MSI}.

The syllabi were coded using the coding scheme in the Table \ref{tab:spinposition} below, which was developed from the literature \cite{Passante, Sadaghiani}. For example, in order for a course to be considered a spin first course, an introduction to the SGE needed to come before introducing the Schr{\"o}dinger Equation. If the syllabus did not provide a course outline for lecture topics, the table of contents of the textbook listed was used to determine which topics come first. 

\section{Results}
\begin{figure}
    \centering
    \includegraphics[width=200 pt, height=100 pt]{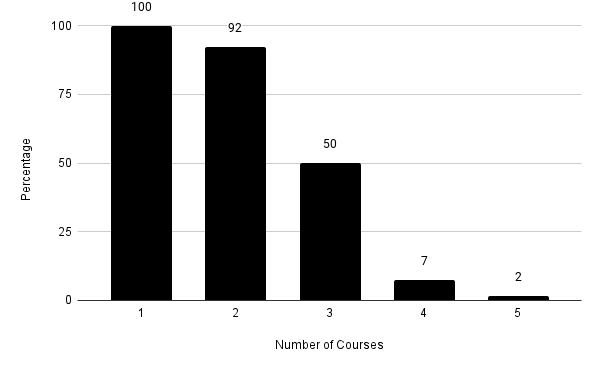}
    \caption{The number of required quantum related courses}
    \label{fig:1}
\end{figure}

From the 188 institutions offering a four year degree in physics, the results of the required number of quantum related courses is shown in Figure \ref{fig:1}. It is shown that all institutions require at least one course, and the large majority (92.0\%) require 2 courses. Exactly half require a third course. The outliers in this data set are the institutions that require four (7.0\%) or five (2.0\%) courses. Of the 8\% of institutions that require only one course, 87\% were modern physics and only 13\% were specifically a quantum mechanics course. 

\begin{figure}
    \centering
    \includegraphics[width=200 pt, height=100 pt]{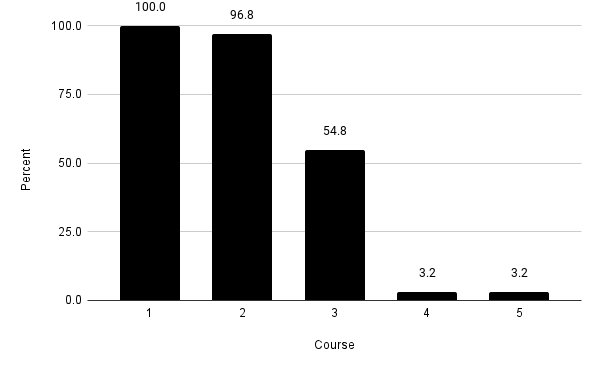}
    \caption{The number of required quantum related courses at MSIs}
    \label{fig:2}
\end{figure}

As stated in the methods, the MSIs were also analyzed separately to compare to the larger data set. The number of required quantum related courses at the MSIs is presented in Figure \ref{fig:2}. Comparing Figures \ref{fig:1} and \ref{fig:2}, it can be concluded that the same trend in number of courses is present in both MSIs and non-MSIs. 

\begin{figure*}
    \centering
    \includegraphics[width=400 pt, height=300 pt]{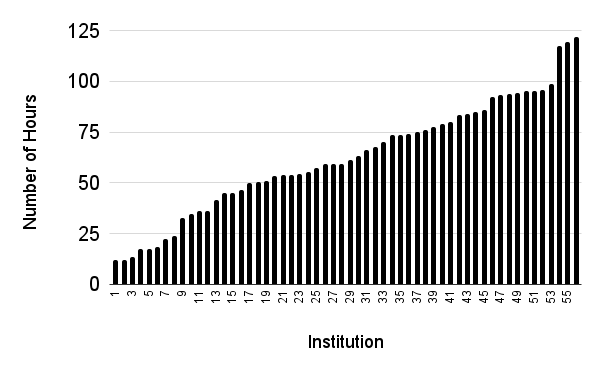}
    \caption{Number of lecture hours on quantum concepts required for each individual institution}
    \label{fig:3}
     \includegraphics[width=400 pt, height=300 pt]{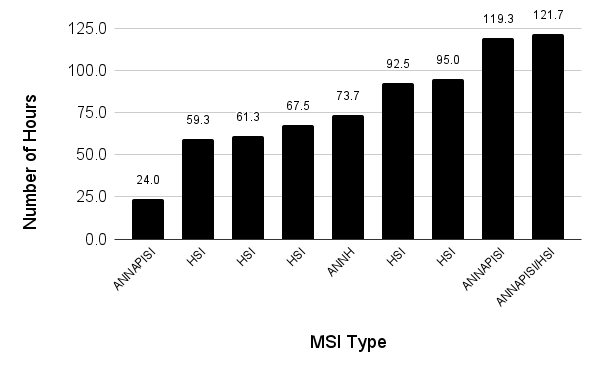}
    \caption{Number of lecture hours on quantum concepts required for each individual MSI}
    \label{fig:4}
\end{figure*}

\begin{figure*}
    \centering
    \includegraphics[width=400 pt, height=300 pt]{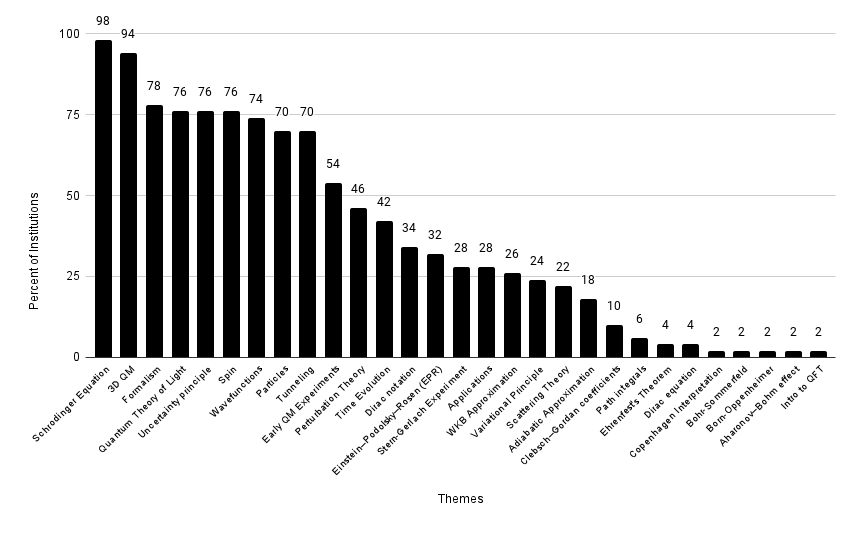}
    \caption{Percentage of themes taught within 50 institutions' quantum curriculum}
    \label{fig:5}
\end{figure*}

From the 56 institutions that syllabi were collected for their entire quantum curriculum, the total lecture hours spent on quantum concepts are presented in Figure \ref{fig:3}. The mean of the 56 institutions is 63.5 hours with a standard deviation of 28.1 hours. The median was 60.3 hours, and the mode was 53.8 hours. The minimum was 12 hours, while the maximum was 121.7 hours. 

The MSIs were again analyzed separately, and their results are presented in Figure \ref{fig:4}. From these 9 institutions, the mean was 79.4 hours with a standard deviation of 31.2 hours. The median 73.7 hours. The minimum was 24.0 hours while the maximum was 121.7 hours. Comparing Figures \ref{fig:3} and \ref{fig:4}, it is shown that the MSIs have a higher mean than the larger data by 15.9 hours, suggesting there is not a disparity present in this data set. 

From the 50 institutions that syllabi were analyzed for themes, it was found that the Schr{\"o}dinger equation (98\%) and three-dimensional quantum mechanics (94\%) were the most commonly found themes across the 50 institutions (Figure \ref{fig:5}. Interestingly, the SGE was only found in 28\% of the curricula, despite research showing its benefit for students learning to discern between physical space and Hilbert space \cite{ZhuSingh2009}.

\begin{figure}
    \centering
    \includegraphics[width=150 pt, height=100 pt]{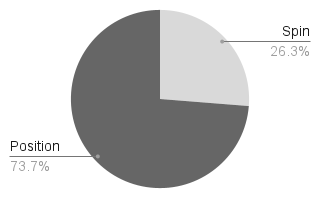}
    \caption{Spin-first versus position-first paradigm used in the quantum mechanics courses of 99 institutions}
    \label{fig:6}
\end{figure}

Of the 112 modern physics or equivalent courses, \textbf{all} used a position-first approach. For the quantum mechanics courses, it was found that spin-first approaches were used by some instructors (26.3\%), but position-first approaches were still the more commonly used approach (73.7\%). 

\begin{figure*}
    \centering
    \includegraphics[width=400 pt, height=300 pt]{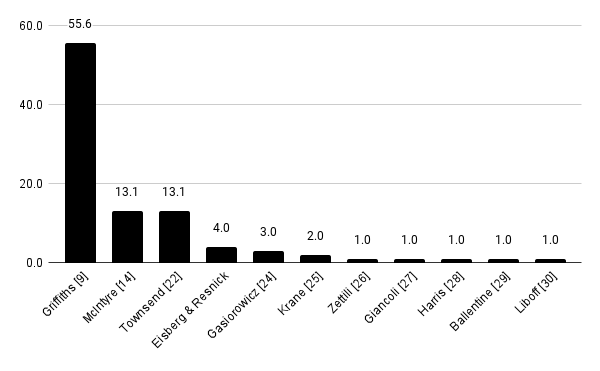}
    \caption{Textbooks used by quantum mechanics courses at 99 institutions
}
    \label{fig:7}
\end{figure*}

\begin{figure}
    \centering
    \includegraphics[width=200 pt, height=200 pt]{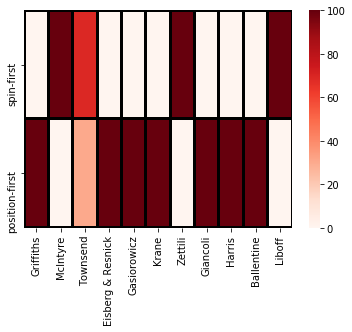}
    \caption{Paradigm used with each textbook}
    \label{fig:8}
\end{figure}

The textbooks used by the 99 quantum mechanics courses were also coded for, with results presented in Figure \ref{fig:7}. Griffiths' textbook \cite{Griffiths} is most commonly used (55.6\%). Both McIntyre \cite{McIntyre} and Townsend \cite{Townsend} were used by 13.1\% of institutions. Eisberg \& Resnick \cite{Eisberg} was used 4\% of the time, Gasiorowicz \cite{Gasiorowicz} 3\%, and Krane \cite{Krane} 2\%. Zettili \cite{Zettili}, Giancoli \cite{Giancoli}, Harris \cite{Harris}, Ballentine \cite{Ballentine}, and Liboff \cite{Liboff} were all used 1\% of the time. A heatmap is presented in Figure \ref{fig:8} that shows the paradigm used in correlation with each textbook.

\section{Discussion}
Across the data set, differences in access and opportunity to learn quantum mechanics were seen. From the results, it has been demonstrated that the range of expected quantum related courses and required quantum course time is quite large. While some students are only expected to enroll in one quantum related course and be exposed to as little as 12 hours of course time, other students are expected to enroll in five courses and experience 121.7 hours of course time.

One interesting finding of this study, although not surprising, is that no modern physics courses are using a spin-first approach. This was expected due to the mathematical preparation of students enrolled in modern physics courses. However, this finding illuminates two key points: (1) An institution might face challenges in using a spin-first approach in their quantum mechanics course if students have previously been exposed to a position-first approach in their modern physics course, and (2) these approaches may have implications for the entire curriculum, not just at the level of individual courses. Future studies should determine if these position-first approaches in modern physics are affecting student learning in a spin-first approach quantum mechanics course.

Despite the physics education community's research and discussions around the spin-first versus position-first approach \cite{Passante, Sadaghiani}, most quantum mechanics instructors are still using a position-first approach. It was also demonstrated that the large emphasis of these courses is placed on the Schr{\"o}dinger equation and its solutions for different potentials, as is common in position-first paradigms \cite{Passante}. However, an interesting finding is that even though instructors are opting for the traditional position-first approach, a very low percentage (28\%) are including the SGE in their course outline. Even if instructors are hesitant to take a spin-first approach, they could incorporate the SGE into their position-first course to help remedy the common difficulty in discerning position space versus Hilbert space \cite{ZhuSingh2009}.

From Figure \ref{fig:8}, one interesting finding is that both spin-first (69.2\%) and position-first (30.8\%) paradigms are used with Townsend \cite{Townsend}, while in all other textbooks only one paradigm was used alongside each. Chapter 1 of Townsend is The Stern-Gerlach Experiment \cite{Townsend}, lending this book to be more readily used in a spin-first approach \cite{Sadaghiani}. This would imply that instructors using this textbook and electing for a position-first approach must be reading chapters out of order in their courses. This may also signify that not all instructors know of the benefits of teaching a spin-first approach \cite{Passante, Sadaghiani, Singh2007185}. Future studies learning instructors' perspectives on the two paradigms could be useful in determining instructors' hesitancies to transform their courses. 

\section{Conclusion}
The current quantum curriculum at research intensive institutions in the US varies widely. Efforts need to be made to decrease this variability and narrow this lack of access for students. Despite the physics education research communities' efforts to promote a spin-first approach highlighting the SGE, most instructors still utilize a position-first approach and place a large emphasis on solving the time-independent Schr{\"o}dinger equation. 

\bibliography{references}

\end{document}